\documentclass[a4paper]{jpconf}
\usepackage{graphicx}
\usepackage[numbers,square,sort&compress]{natbib}
\begin{document}
	
\title{Investigation of the magnetic ground state of PrRu$_2$Ga$_8$ compound}

\author{Michael O Ogunbunmi and Andr\'{e} M Strydom$^{*}$}

\address{Highly Correlated Matter Research Group, Physics Department, University of Johannesburg, P. O. Box 524,  Auckland Park 2006, South Africa.}

\ead{amstrydom@uj.ac.za}
\begin{abstract}
	We have investigated the ground state properties of the orthorhombic structure compound PrRu$_2$Ga$_8$ through electronic and magnetic properties studies. The compound crystallizes in the CaCo$_2$Al$_8$-type structure, belonging to space group $Pbam$ (No. 55). The temperature dependence specific heat shows a $\lambda$-type anomaly at $T_N$ = 3.3 K, indicating a bulk phase transition probably of antiferromagnetic origin. At the N\'{e}el temperature $T_N$, the entropy approaches the value of 4.66~J/mol.K which is about 0.8Rln(2), where R is the universal gas constant. The analysis of the low temperature specific heat gives $\gamma$ = 46 mJ/mol.K$^2$. The temperature dependence DC magnetic susceptibility $\chi(T)$ confirms the anomaly at 3.3 K and follows the Curie-Weiss law for temperatures above 50~K, with the calculated effective magnetic moment, $\mu_\mathrm{{eff}}$ = 3.47(2)~$\mu_B$/Pr and Weiss temperature $\theta_p$ = --7.80(1)~K. This effective magnetic moment value is in good agreement with the Hund$^\prime$s rule theoretical free-ion value of 3.58~$\mu_B$ for Pr$^{3+}$. The electrical resistivity data also shows an anomaly at $T_N$ and a broad curvature at intermediate temperatures probably due to crystalline electric field (CEF) effects. The Pr$^{3+}$ in this structure type has a site symmetry of $C_s$ which predicts a CEF splitting of the $J$ = 4 multiplet into 9 singlets and thus rule out in principle the occurrence of spontaneous magnetic order. In this article we discuss the magnetic order in PrRu$_2$Ga$_8$  in line with an induced type of magnetism resulting from the admixture of the lowest CEF level with the first excited state. 
	
\end{abstract}

\section{Introduction}
The Pr$T_2X_8$ [$T$ = Fe, Co, Ru, Rh; $X$=Al, Ga, In] family of compounds are quasi-skutterudites which crystallize in the orthorhombic CaCo$_2$Al$_8$-type structure with $Pbam$ space group (No. 55) \cite{schluter2001ternary}. The structure of this family of compounds consists of caged network of atoms and belongs to a class of largely unexplored rare-earth intermetallic compounds. The Pr and $T$ atoms in this structure form a chain parallel to the $c$-axis and are both separated by the Ga atoms. The Pr$^{3+}$ ion has a site symmetry of monoclinic $C_s$ in the crystal structure and as a consequence, the CEF splitting of the $J$= 4 multiplet results in 9 singlets \cite{henderson2005crystal,nair2016magnetic}. In such systems, the ground state is expected to be a $\Gamma_1$ singlet as dictated by Pr$^{3+}$ site symmetry. Hence, the occurrence of spontaneous magnetic order in this family especially where the rare-earth site symmetry predicts a singlet ground state is unexpected. However, recent experimental studies have revealed that in exceptional cases, systems predicted to have a $\Gamma_1$ singlet ground state eventually show a magnetic ordering at low temperatures \cite{nair2016magnetic,adroja2012inelastic,anand2014investigations,schobinger2001magnetic,blanco1992metamagnetism,andres1972induced,nair2017pr,vejpravova2007magnetic}. Magnetism in these systems are largely attributed to induced moment magnetism due to the overcritical exchange interactions between the ground state singlet and the first excited state. In this article, we present results of our investigation on the nature of the ground state in PrRu$_2$Ga$_8$ compound, the crystal structure of which was first announced by Schl{\"u}ter and Jeitschko \cite{schluter2001ternary}.

\section{Experimental methods and crystal structure}
\begin{figure}[!h]
	\centering
	\includegraphics[scale=0.48]{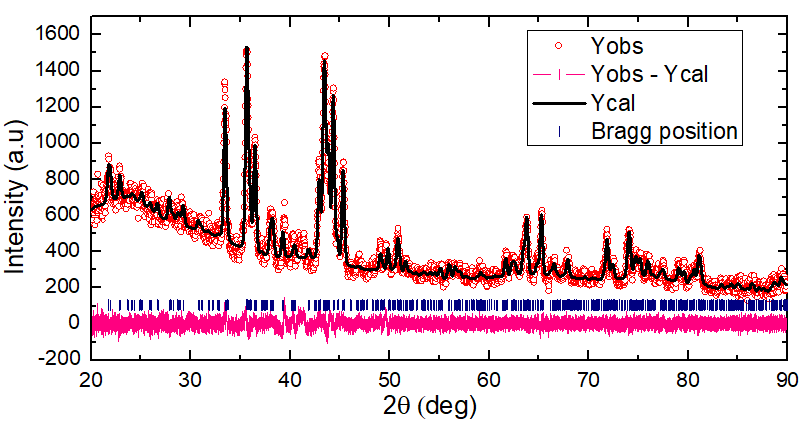}
	\caption{Experimentally observed X-ray diffraction pattern of PrRu$_2$Ga$_8$ (red circle) along with a Rietveld refinement profile (black line) based on the $Pbam$ space group. The Bragg peaks are shown as blue vertical bars.\label{structure}}
\end{figure}

Polycrystalline samples of PrRu$_2$Ga$_8$ and LaRu$_2$Ga$_8$ were prepared by arc melting stoichiometric amounts of Pr, La, Ru and Ga (4N) in an Edmund B{\"u}hler arc furnace by a method described in ref.~\cite{nair2016magnetic}. Room temperature powder X-ray diffraction (XRD) was recorded using a Rigaku Smartlab diffractometer with Cu-K$_\alpha$ radiation. The observed pattern for both samples correctly match that of the $Pbam$ space group and no impurity phases were observed within the resolution limit of the instrument. A Rietveld refinement \cite{thompson1987rietveld} using Fullprof prgram \cite{rodriguez1990fullprof} was carried out on the patterns collected and the lattice parameters obtained are presented in Table~\ref{PrT2Ga8_structural_parameters} which are consistent with earlier reports \cite{schluter2001ternary,nair2016magnetic,tougait2005prco}. 
\begin{table}[t!]
	\centering

	\caption{Crystallographic parameters of  PrRu$_2$Ga$_8$ and LaRu$_2$Ga$_8$ obtained from the Rietveld refinements.}
	\label{PrT2Ga8_structural_parameters}
	\begin{tabular}{|c|c|c|}
		\hline
		Compound& PrRu$_2$Ga$_8$  & LaRu$_2$Ga$_8$ \\ 
		Space group& $Pbam$ (No. 55)  & $Pbam$ (No. 55)\\ 
		$a$ (\AA)& 12.607(3) & 12.65(3)5\\
		$b$ (\AA)& 14.713(3) & 14.715(5)\\
		$c$ (\AA) &  4.1010(2) & 4.112(2) \\
		$V$ (\AA$^3$) & 760.68 (4)&766.90 (1)\\
		$R_p (\%)$ &6.443 & 5.610 \\
		$R_{wp} (\%)$ & 8.392& 8.237 \\
		\hline

	\end{tabular}

\end{table}
The Rietveld refinement of the XRD pattern of PrRu$_2$Ga$_8$ is shown in Fig.~\ref{structure}. In this structure, the shortest Pr-Pr separation is 4.080~\AA~(about 26\% greater than the sum of two Pr ionic radii), Pr-Ga is 3.123~\AA~ (about 7\% greater than the sum of their ionic radii)~ and Pr-Ru is 3.101~\AA~ (about 7\% greater than the sum of their ionic radii). In view of this,  the structure of PrRu$_2$Ga$_8$ therefore resembles those of other caged compounds where the Pr atom is enclosed in an oversized cage network formed by Ru-Ga atoms.  As stated above, the shortest Pr-Pr separation in the structure which is $\sim$ 26\% is greater than the separation between the sum of two Pr ionic radii suggests that the Pr atoms are weakly bonding or may be out of reach of direct magnetic exchange due to no direct orbital overlap. A possible consequence of this will be the suppression of the magnetic transition temperature in systems with a magnetic order parameter in the ground state. \\
Magnetic properties have been measured using the Magnetic Property Measurement System (Quantum Design Inc. San Diego) between 1.9~K and 300~K with an external magnetic field up to 7~T. The electrical resistivity measurement from 300~K down to 1.9~K was taken using the conventional four probe DC method with contacts made using  a spot welding equipment.  Specific heat was measured using the quasi-adiabatic thermal relaxation method down to 0.4~K. Both the electrical resistivity and specific heat were measured using the Physical Property Measurement System also from Quantum Design.\\

\begin{figure}[!t]
	\centering
	\includegraphics[scale=0.55]{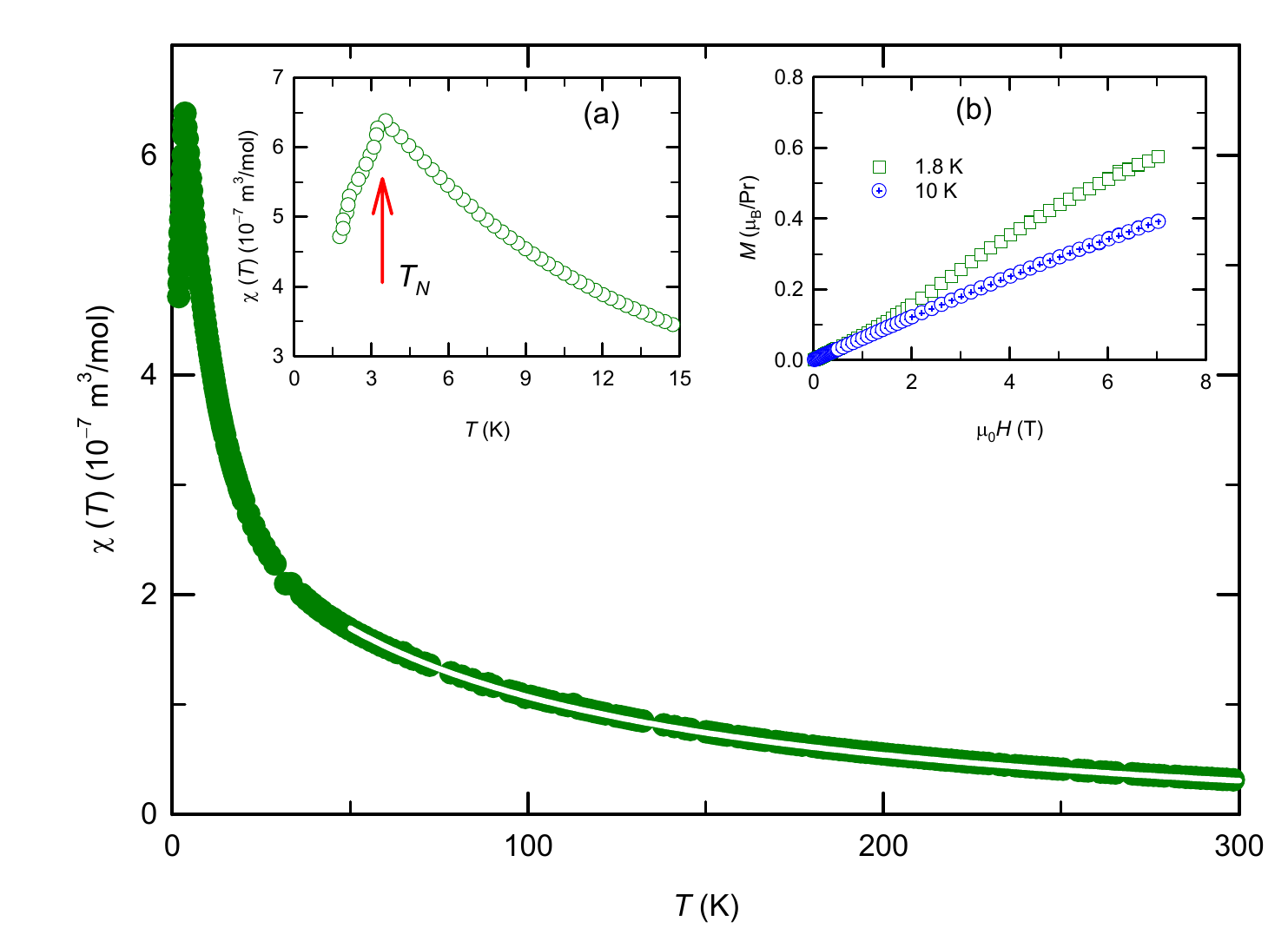}
	\caption{Temperature dependence of magnetic susceptibility $\chi(T)$ of PrRu$_2$Ga$_8$ together with a Curie-Weiss fit (white-solid line). Inset (a): Low-$T$ of $\chi(T)$ showing an antiferromagnetic ordering at $T_N$ = 3.3~K. (b): Isothermal magnetization of PrRu$_2$Ga$_8$ at 1.8~K and 10~K.\label{magnetic}}
\end{figure}
\section{Magnetic properties}
The temperature dependence of magnetic susceptibility $\chi(T)$ of PrRu$_2$Ga$_8$ measured between 1.9 and 300~K is presented in Fig.~\ref{magnetic}. For temperatures above 50~K, $\chi(T)$ could be fitted to the Curie-Weiss expression given by; $\chi = N_A \mu_\mathrm{eff}^2/(3k_B(T-\theta_p))$,
where $\mu_\mathrm{eff}$ and $\theta_p$ are the effective magnetic moment and Weiss temperature respectively, N$_A$ is the Avogadro's number and k$_B$ is the Boltzmann's constant. Values of $\mu_\mathrm{{eff}}$ = 3.47(2)~$\mu_B$  and $\theta_p$ = $-$7.80(1)~K are obtained from the least-squares fit. The value of $\mu_\mathrm{{eff}}$ obtained is fairly reconcilable to the value of $g_J\sqrt{J(J+1)}\mu_B = 3.58~\mu_B$ expected for a free Pr$^{3+}$ ion. At $\sim$ 3.3~K, $\chi(T)$ develops an anomaly signaling a phase transition possibly of antiferromagnetic origin. The low temperature region of $\chi(T)$ is expanded in inset (a) with the magnetic transition at $T_N$ indicated by the arrow. In inset (b), the isothermal magnetization $M(B)$ of PrRu$_2$Ga$_8$ is presented in fields up to 7~T. At 1.8~K, an upward curvature in $M(B)$ is seen near 2~T, while closer to 7~T the curvature turns slightly downward again mimicking saturation. The origin of these features are not immediately clear but appears to be matamagnetic in nature. The behaviour of $M(B)$ in 10~K however follows a quasi-linear dependence with  fields.

\section{Specific heat}
\begin{figure}[!t]
	\centering
	\includegraphics[scale=0.55]{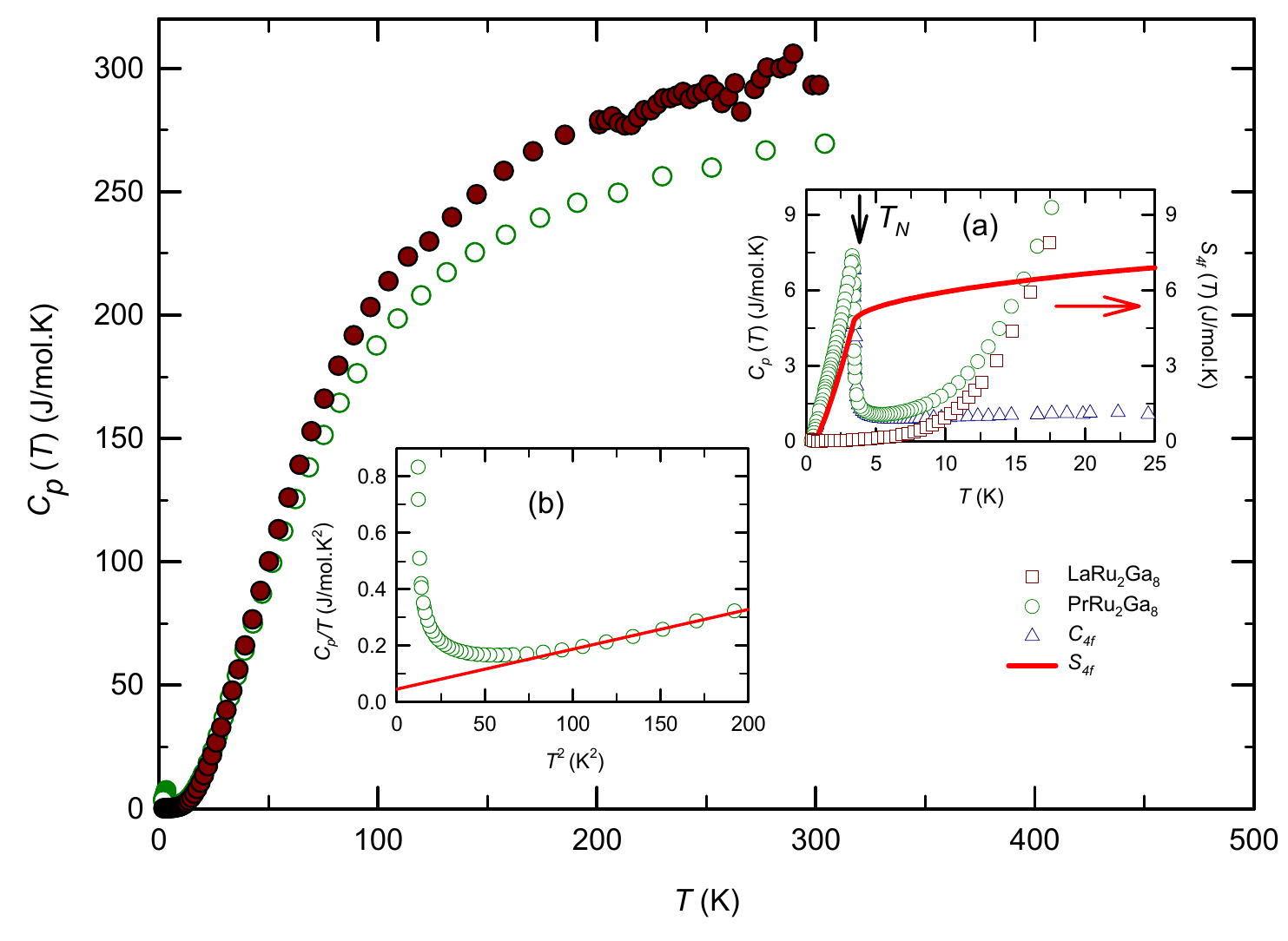}
	\caption{Temperature dependence of specific heat $C_p(T)$ of  PrRu$_2$Ga$_8$ and  LaRu$_2$Ga$_8$. Inset (a): Low-$T$ $C_p(T)$ of  PrRu$_2$Ga$_8$ and LaRu$_2$Ga$_8$, electronic contribution to specific heat $C_{4f}(T)$ and  the magnetic entropy $S_{4f}(T)$. (b): Plot of $C_p/T$ against $T^2$ together with a Debye fit (red-solid line). \label{specific}}
\end{figure}

The temperature dependence of specific heat $C_p(T)$ of PrRu$_2$Ga$_8$ and that of the nonmagnetic reference compound LaRu$_2$Ga$_8$ is presented in Fig.~\ref{specific}. In inset (a), the low temperature $C_p(T)$ of PrRu$_2$Ga$_8$ and LaRu$_2$Ga$_8$ together with the electronic contribution to specific heat $C_{4f}(T)$ obtained by subtracting the specific heat of LaRu$_2$Ga$_8$ from that of the main compound PrRu$_2$Ga$_8$ and the calculated entropy $S_{4f}(T)$ are shown. A $\lambda$-type anomaly at $T_N$ $\approx$ 3.3~K is clearly indicated by the arrow in the figure for both PrRu$_2$Ga$_8$ and $C_{4f}(T)$ data indicating a bulk phase transition. $S_{4f}(T)$ is calculated using the expression; $S_{4f}(T^\prime) = \int_{0}^{T^\prime}C_{4f}(T)/T dT$.
%
The value of the entropy around $T_N$ is 4.66 J/mol.K which is $\simeq$ 80\% of Rln(2) expected for a doublet ground state. We note however that the full doublet entropy is only released at about 8 K. The estimation of $C_{4f}(T)$ was only carried out up to about 24~K, above which the phonon specific heat dominates. Within this temperature range, a plot of $C_{4f}(T)/T$ against $T^2$  based on the expression $C_{4f}(T)/T(T\longrightarrow$0) $\equiv$ $\gamma$, was found not to be an increasing function of $T$ which consequently renders the extraction of the electronic Sommerfeld coefficient, $\gamma$ from $C_{4f}(T)$ difficult in the present case. We have therefore used the total specific heat in the estimation of $\gamma$ as presented in inset (b) of Fig. \ref{specific} based on the expression $C_p/T = \gamma + \beta T^2$ and $\beta = 12\pi^4nR/(5\theta^3_D$), where $n$ and R are the number of atoms and universal gas constant respectively.
%
From the least-squares fit, a value of $\gamma$ = 46.04(3)~mJ/mol.K$^2$ and Debye temperature $\theta_D$ = 248.1(2)~K are obtained. The $\gamma$ value obtained here is slightly enhanced compared to that found for ordinary metals which could be due to moderate heavy-electron-like behaviour in the system.
The Sommerfeld coefficient $\gamma$ has a direct relationship with the mass of the quasiparticles in a metal at low temperatures and it gives an idea about their degree of mass enhancement. 
We note that in a similar iso-structural compound PrCo$_2$Ga$_8$ \cite{Ogunbunmi2017electronic}, an enhanced quasi-particle mass behaviour has been observed while on the other hand, iso-structural aluminides compounds like PrFe$_2$Al$_8$ and PrCo$_2$Al$_8$ \cite{nair2016magnetic,tougait2005prco} have been reported to have $\gamma$ values in the range of ordinary metals ($\approx$ 10~mJ/mol.K$^2$). 
\begin{figure}[!h]
	\centering
	\includegraphics[scale=0.55]{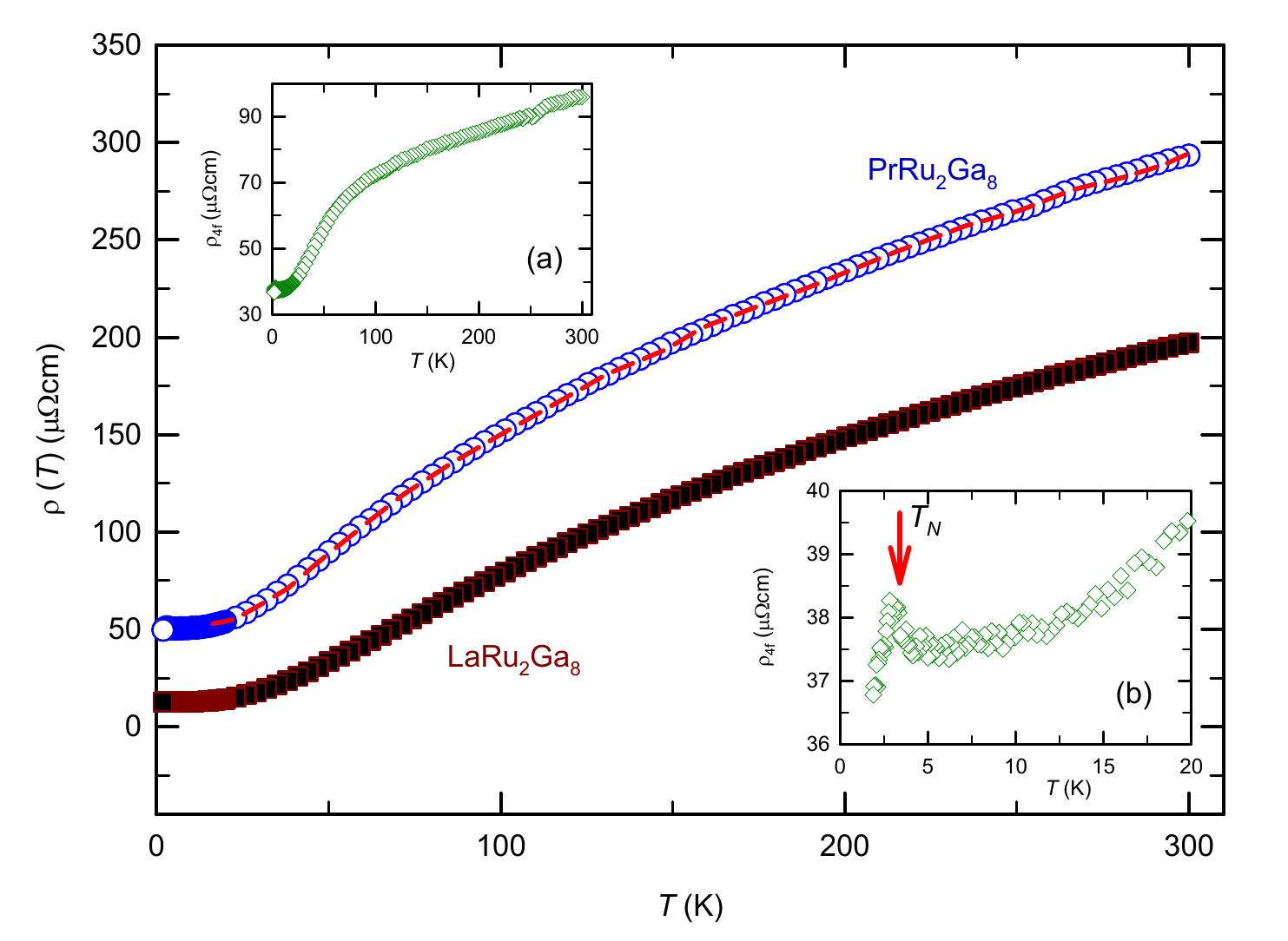}
	\caption{Temperature dependence of electrical resistivity $\rho(T)$ of PrRu$_2$Ga$_8$ and LaRu$_2$Ga$_8$. The red-dashed line is a BGM fit described in the text.  Inset (a):  Electronic contribution to resistivity $\rho_{4f}(T)$. (b):  Low-$T$ of  $\rho_{4f}(T)$ with an arrow indicating $T_N$ = 3.3~K. \label{resistivity1}}
\end{figure}

\section{Electrical Resistivity}
The temperature dependence of electrical resistivity $\rho(T)$ of PrRu$_2$Ga$_8$ and LaRu$_2$Ga$_8$ measured between 1.9 and 300~K is presented in Fig.~\ref{resistivity1}. LaRu$_2$Ga$_8$ shows a typical metallic behaviour from room temperature down to low temperatures with residual resistivity ratio (RRR) of 15.47. $\rho(T)$ of PrRu$_2$Ga$_8$ shows a broad curvature at intermediate temperatures and an anomaly indicating a phase transition at $T_N$ $\approx$ 3.3~K. The RRR of $\approx$ 6 is observed which is lower compared to that of LaRu$_2$Ga$_8$. However, the RRR of both compounds suggest a reasonable good crystalline quality. Furthermore, for temperatures above 10~K, $\rho(T)$ of PrRu$_2$Ga$_8$ was fitted to the Bloch-Gr\"{u}neissen-Mott (BGM) model \cite{mott1958theory} as indicated by the red-dashed line. The BGM expression is given by;
\begin{equation}
\rho(T) = \rho_0 + \frac{4K}{\Theta_D}\left(\frac{T}{\Theta_D}\right)^5\int_{0}^{\Theta_D/T} \frac{x^5 dx}{(e^x -1)(1-e^{-x})} + \alpha T^3,
\label{BGM}
\end{equation}	
where $\rho_0$ is the residual resistivity due to defect scattering in the crystal lattice, $K$ is the electron-phonon coupling constant, $\Theta_D$ is the Debye temperature and also contains a contribution from the electron-electron correlations \cite{falkowski2017new,pikul2003kondo} while $\alpha T^3$ is the Mott term which describes the $s$-$d$ interband scattering. Values of $\rho_0$ =  49.522~$\mu \Omega$ cm, $K$ =  67.588~$\mu \Omega$ cm K, $\Theta_D$ = 41.162~K and $\alpha$ =  $-$2.607 $\times$ 10$^{-6}$~$\mu \Omega$ cm K$^{-3}$ are obtained from the fit. The electronic contribution to resistivity $\rho_\mathrm{4f}$ obtained by subtracting the phonon contribution from PrRu$_2$Ga$_8$ is presented in inset (a) while (b) is the low temperature plot of $\rho_\mathrm{4f}$. $\rho_\mathrm{4f}$ shows a strong temperature dependence from room temperature down to $\approx$ 30~K with a shallow curvature around 100~K. This feature likely originates from possible CEF effect on the resistivity. Below about 30~K, the resistivity is weakly temperature dependent down to $\approx$ 5~K, below which it rises into a peak centred at 3.3~K which is associated with the magnetic ordering at $T_N$.



\section{Discussion and conclusion}
From the electronic and magnetic properties of PrRu$_2$Ga$_8$, a phase transition at $T_N$= 3.3~K is observed. The magnetic ordering in PrRu$_2$Ga$_8$ is at variance to the singlet ground state expected based on the $C_s$ site symmetry of the Pr$^{3+}$ in the CaCo$_2$Al$_8$ structure type. Among other factors, induced magnetism is thought to have been responsible for such observation arising due to the admixture of the first excited CEF level with the ground state singlet when the exchange interaction exceeds a critical value based on the expression for self-induced moment ordering \cite{adroja2012inelastic,anand2014investigations};
\begin{equation}
T_c = \Delta\left[\ln\frac{J_{ex}\alpha^2 + n\Delta}{J_{ex}\alpha^2 + n\Delta}\right]^{-1},
\end{equation}
where $\Delta$ is the energy splitting between the ground state singlet and the first excited state, $\alpha$  is the matrix element between the ground state singlet and the first excited state, $n$ is the degeneracy of the first excited state, $J_{ex}$ is the exchange interaction and $T_c$ is the mean-field critical temperature. From inelastic neutron scattering (INS) experiment, the exchange interaction can therefore be estimated based on the above expression. In the present analysis, the exchange interaction can be predicted according to the expression; $\theta_p = -J_{ex} J(J+1)/{3k_B}$,
%
where $\theta_p$ = $-$7.8~K as obtained from magnetic susceptibility analysis, and $J$ = 4. The value of $J_{ex}$ = 0.094~meV is obtained which is of the same order magnitude with the ordering temperature found in PrRu$_2$Ga$_8$. An induced magnetic ordering is thus believed to be favoured when the value of 0.094~meV is comparable or greater than that estimated from the INS experiment. A $\gamma$ value of 46~mJ/mol.K$^2$ indicates a possible moderate heavy-electron like behaviour in the material.
The magnetic susceptibility follows the Curie-Weiss law for temperatures above 50~K and gives $\mu_\mathrm{{eff}}$ = 3.47~$\mu_B$/Pr  which is close to that expected for a free Pr$^{3+}$ ion. Magnetic phenomena  in this system may therefore be expected to be governed by a stable and well-defined Pr local magnetic moment. Further measurements in fields and inelastic neutron scattering are required to further explore microscopic aspects and crystal field effects discussed in this article.

\section*{Acknowledgement}
MOO acknowledges the UJ-URC bursary for doctoral studies in the Faculty of Science. AMS thanks the SA-NRF (93549) and UJ-URC for financial support.

\bibliographystyle{iopart-num}
\bibliography{PrRu2Ga8}


\end{document}